\documentclass[12pt]{iopart}
\usepackage{iopams}
\usepackage{graphicx}  
\usepackage{wrapfig}

\def\Cref{C_{\rm ref}}

\def\grho1{\left<|\nabla{\rho}|\right>}
\def\grhov1{\left<|\nabla{\rvol}|\right>}
\def\Bpol{B_{\theta}}

\def\rvol{\rho_{\rm V}}

\def\q95{q_{95}}

\providecommand{\Cref}[1]{Chapter~\ref{#1}}
\providecommand{\cref}[1]{chapter~\ref{#1}}

\def\a0v{a_{\rm 0,V}}
\def\b0v{\beta_{\rm 0,V}}

\def\grho{\left \langle \left \vert \nabla\rho \right\vert\right\rangle}

\def\grho1{\left<|\nabla{\rho}|\right>}
\def\grhov1{\left<|\nabla{\rvol}|\right>}
\def\Bpol{B_{\theta}}

\def\rvol{\rho_{\rm V}}

\def\q95{q_{95}}
\def\ur0{u_{r,0}}
\def\upol0{u_{\theta,0}}
\def\uphi0{u_{\phi,0}}
\def\Bpol0{B_{\theta,0}}
\def\Bphi0{B_{\phi,0}}
\def\urt{\tilde{u_{r}}}

\def\upolt{\tilde{u_{\theta}}}
\def\uphit{\tilde{u_{\phi}}}
\def\Brhot{\tilde{B_{r}}}
\def\Bpolt{\tilde{B_{\theta}}}
\def\Jpolt{\tilde{J_{\theta}}}
\def\Bphit{\tilde{B_{\phi}}}

\def\chihat{\hat{\chi}}
\providecommand{\Cref}[1]{Chapter~\ref{#1}}
\providecommand{\cref}[1]{chapter~\ref{#1}}

\begin{document}
\title[Flux pumping transport mechanisms]{Investigating the mechanisms behind magnetic and kinetic transport in flux pumping scenarios: linear and early saturation phases}

\author{E. Fable, H. Zhang, F. Stefanelli, H. Zohm}


\address{Max-Planck-Institut f\"ur Plasmaphysik, 85748 Garching, Germany}
\ead{emiliano.fable@ipp.mpg.de}

\begin{abstract}
The physics mechanisms behind the magnetic field dynamo, energy and particle transport observed during a magneto--hydro--dynamic (MHD) flux pumping scenario are investigated analytically in the context of one--fluid full MHD equations and compared to the results obtained from JOREK simulations.
Some of the mechanisms detailed maybe present in general during saturated core MHD activity characterized by plasma flow perturbations with a finite toroidal wavenumber. The focus in this work is on the linear and early saturation phase. The long--time non--linear dynamics will be investigated in a following work.
\end{abstract}

\maketitle

\section{Introduction}
The core of toroidal magnetized plasma is subjected by intense heating and thus  chracterized by substantial peaking of the pressure profile. This in turn can, in conjunction with the background current density profile peaking, drive global internal MHD modes, of which the 1/1 mode (toroidal and poloidal mode numbers) is of interest in this work \cite{bussacinternalkinkprl}.
The growing helical structure pushes the plasma against the outermost unstable resonant surface responsible for the mode (e.g. $q=1$ for a sawtooth crash \cite{sawtoothpaperkad1,sawtoothpaperjet1,sawtoothpaperporcelli}), resulting in either sudden collapse of the core pressure via reconnection, or saturation to a level that does not completely collapse the pressure \cite{kinksaturatedbrunetti}. 
In this second scenario, the appearance of a continous magnetic dynamo that keeps the current profile from peaking again \cite{jardinprlfluxp,krebsfluxp} can maintain the saturated mode alive but at the same time acting also on the pressure and density profiles \cite{Jorekhaoweifp1}. 

To be able to develop a reduced model of this phenomenon for integrated modeling frameworks such as ASTRA8 (A8) \cite{astracode1,astracode2,astracode3}, one first has to understand the mechanisms behind the observed magnetic dynamo and the core pressure regulation arising from the saturated MHD mode (whereas the role of more short Larmor-scale instabilities will be investigated elsewhere). In this direction we consider a perturbation non--linear analysis framework. Field and flow perturbations couple non--linearly and this are the objects that ultimately lead to transport of macroscopic profiles (in the sense of not only slow but also axisymmetric dynamics). It is observed that these mechanisms are somewhat independent of the specific MHD mode causing them, as such some of the conclusions reached here in the context of flux pumping may be applicable to the internal kink as well, which is responsible for the sawooth crash, or the internal tearing mode \cite{sawdynamopopcollam2026}.

As mentioned in the abstract, this paper is the first of a two--parters to address the dynamics and transport caused by the instability over the three phases: linear, saturation, and long--time behavior. In this work we address the first two, whereas the long--time dynamics will be addressed in the next second part work.

The present paper is organized as follows: Section 2 presents the model equations and the split in equilibrium and perturbation quantities. Section 3 goes deeper in the analysis of perturbations coupling and the effect on macroscopic profiles. Section 4 presents the comparison to JOREK simulations. Section 5 draws conclusions and an outlook for further studies.
\section{MHD model equations}
The starting point is the same one--fluid full MHD framework considered in the JOREK code \cite{Jorekhoelzl}:
\begin{eqnarray}
\partial_t \mathbf{A}=-\nabla\Phi + \mathbf{u}\times \mathbf{B}- \eta R J_\phi \nabla\phi \nonumber \\
\rho_M \partial_t \mathbf{u} = -\nabla P + \mathbf{J}\times \mathbf{B} + \mu \nabla^2 \mathbf{u} \nonumber \\
\partial_t P = -\mathbf{u}\cdot\nabla P -\frac{5}{3} P \nabla\cdot \mathbf{u}  + \frac{2}{3}\left(Tr + \mathcal{P}\right)
\label{jorekmhdeq1}
\end{eqnarray}
where $\mathbf{A},\Phi,\mathbf{u},P$ are respectively the vector potential, the electrostatic potential, the plasma flow, and the pressure fields. All quantities are in SI units except fields that are linked to time or speed are normalized by the unitary (in field strength) Alfv\'en speed $V_a = 1/\sqrt{\mu_0 \rho_0}$ with $\mu_0$ the vacuum magnetic permeability and $\rho_0$ a dimensional plasma mass density of reference. Currents $\mathbf{J}$ are normalized to $\mu_0$. Auxiliary fields are the magnetic field $\mathbf{B}=\nabla\times\mathbf{A}$ and the plasma current density $\mu_0 \mathbf{J}=\nabla\times\mathbf{B}$. $\phi$ is the toroidal angle coordinate, $\eta$ the resistivity, $\mu$ the viscosity, $\rho_M$ the plasma mass density. $Tr,\mathcal{P}$ are the energy transport and power source terms respectively. The current drive source is not explicitly written but it can be present, but it makes no relevance to the discussion later on. The gauge choosen here is $\nabla\cdot\mathbf{A}=0$, whereas JOREK works in the gauge $\Phi=0$. The coordinate system is either cylindrical $(R,Z,\phi)$ or flux--surface based $(\rho,\theta,\phi)$. The radial coordinate of choice is $\rho=\sqrt{\Psi/\Psi_b}$ (also called "rho poloidal"), i.e. the normalized poloidal flux under square root. 

Now we split the fields in axisymmetric (Fourier toroidal number $n=0$) and non--axisymmetric ($n\neq 0$) components, and take the toroidal angle average $\langle ... \rangle_\phi=\frac{1}{2\pi}\int_0^{2\pi} ... d\phi$ to obtain for once the averaged fields equations (fields with pedix "0", whereas non--axisymmetric perturbations have a tilde over the letter):
\begin{eqnarray}
\partial_t \mathbf{A_0}=-\nabla\Phi_0 + \langle\mathbf{\tilde{u}}\times \mathbf{\tilde{B}}\rangle_\phi + \mathbf{u_0}\times \mathbf{B_0}- (\eta R J_\phi \nabla\phi)_0 \nonumber \\
\rho_M \partial_t \mathbf{u_0} = -\nabla P_0 + \langle\mathbf{\tilde{J}}\times \mathbf{\tilde{B}}\rangle_\phi + \mathbf{J_0}\times \mathbf{B_0} + \mu \nabla^2 \mathbf{u_0} \nonumber \\
\partial_t P_0 = -\langle\mathbf{\tilde{u}}\cdot\nabla \tilde{P}\rangle_\phi-\mathbf{u_0}\cdot\nabla P_0 -\frac{5}{3} P_0 \nabla\cdot \mathbf{u_0} + \frac{2}{3}\left(
Tr + \mathcal{P}\right)
\label{jorekmhdeqequil1}
\end{eqnarray}
where we have dropped negligible contributions such as the perturbed resistivity as well as the perturbations entering into the terms denoted as $Tr,\mathcal{P}$. While they are relvant to dissipation of the free energy on the long--time scale, they are much smaller than other terms in the equations. We assume dominantly incompressible flows, that is $\nabla\cdot\mathbf{\tilde{u}} \approx 0$.

From these averaged equations one can immediately identify the dynamo term $\langle\mathbf{\tilde{u}}\times \mathbf{\tilde{B}}\rangle_\phi$ and the energy sink/source term due to the mode $-\langle\mathbf{\tilde{u}}\cdot\nabla \tilde{P}\rangle_\phi$. Now it is important to also consider $\mathbf{u_0}\times \mathbf{B_0}$ in the Ohm's law and $-\mathbf{u_0}\cdot\nabla P_0 -\frac{5}{3} P_0 \nabla\cdot \mathbf{u_0}$ in the pressure evolution equation. The flow $\mathbf{u_0}$ there is not only any equilibrium flow pre--existent to the perturbation, but it is also the equilibrium flow {\it generated} by the mode itself, especially any radial (across flux surfaces) component. It will be shown later that these terms are very relevant to the long--time scale saturated phase dynamics.  

The $n\neq 0$ perturbation fields equations are:
\begin{eqnarray}
\fl \partial_t \mathbf{\tilde{A}}=-\nabla\tilde{\Phi} + \widetilde{\mathbf{\tilde{u}}\times \mathbf{\tilde{B}}}+ \mathbf{u_0}\times \mathbf{\tilde{B}}+ \mathbf{\tilde{u}}\times \mathbf{B_0}  \nonumber \\
\fl \rho_M \partial_t \mathbf{\tilde{u}} = -\nabla \tilde{P} + \widetilde{\mathbf{\tilde{J}}\times \mathbf{\tilde{B}}}+ \mathbf{J_0}\times \mathbf{\tilde{B}}+ \mathbf{\tilde{J}}\times \mathbf{B_0} + \mu \nabla^2 \mathbf{\tilde{u}} \nonumber \\
\fl \partial_t \tilde{P} = -\mathbf{\tilde{u}}\cdot\nabla P_0-\mathbf{u_0}\cdot\nabla \tilde{P}-\widetilde{\mathbf{\tilde{u}}\cdot\nabla \tilde{P}} -\frac{5}{3} \tilde{P} \nabla\cdot \mathbf{u_0} 
\label{jorekmhdeqpert1}
\end{eqnarray}
The overtilde $\widetilde{\mathbf{\tilde{f}}\times \mathbf{\tilde{g}}}$ indicates that its toroidal average has been automatically subtracted, since it goes into equations (\ref{jorekmhdeqequil1}).
The non--linear quadratic terms are retained. They can contribute to the saturation mechanism (see later discussion in the simplified model) \cite{rosenbluthnonlinmhdpaper1,paononlinmhdpaper}.

One immediate observation that can be made on the set (\ref{jorekmhdeqequil1}) is that the momentum balance in the radial direction is expected to be dominated by the pressure/Lorentz force balance $\partial_r P_0 \approx (\mathbf{J_0}\times\mathbf{B_0})_r$, as such the radial equilibrium flow $\ur0$ cannot be determined from this balance. Instead, we exploit the poloidal projection of the equilibrum Ohm's law, which results in:
\begin{eqnarray}
0 \approx -\frac{1}{r}\partial_\theta\Phi_0 + (\langle\mathbf{\tilde{u}}\times \mathbf{\tilde{B}}\rangle_\phi)_\theta - \ur0 \Bphi0
\label{polradohmslaweq1}
\end{eqnarray}
The equilibrium electrostatic potential $\Phi_0$ poloidal derivative can be ignored since its neoclassical origin in the very plasma core is negligible compared to the MHD perturbations. As such, we can find the origin of the equilibrium radial flow as:
\begin{eqnarray}
\ur0  \approx \frac{1}{\Bphi0} (\langle\mathbf{\tilde{u}}\times \mathbf{\tilde{B}}\rangle_\phi)_\theta 
\label{radfloworigin1}
\end{eqnarray}
This will be confirmed to be the case later from the JOREK simulations. This flow is crucial to explain both the dynamo and the pressure evolution. In particular since the compression term $-5/3 P_0 \nabla\cdot\mathbf{u_0}$, when taking the flux surface average $\langle ... \rangle_{f.s.a.}=\partial_V \int dV (...)$, reduces to the radial derivative of the radial flow $\ur0$.
Note that the electrostatic potential perturbation $\tilde{\Phi}$ can be computed in the saturation phase as:
\begin{eqnarray}
\tilde{\Phi} = -\int^\theta \left [r \tilde{u}_r {B}_{\phi,0}\right]d\theta
\label{elecpotpert1}
\end{eqnarray}
since the poloidal projecton of linearized Ohm's law is approximately:
\begin{eqnarray}
0 \approx -\tilde{u}_r {B}_{\phi,0} - \frac{1}{r}\partial_\theta\tilde{\Phi}
\label{Ohmslaw2}
\end{eqnarray}
whereas in the radial direction:
\begin{eqnarray}
0 \approx \tilde{u}_\theta {B}_{\phi,0}-\tilde{u}_\phi {B}_{\theta,0} - \partial_r\tilde{\Phi}
\label{Ohmslaw3}
\end{eqnarray}
leading to:
\begin{eqnarray}
\tilde{u}_\phi \frac{{B}_{\theta,0}}{{B}_{\phi,0}} = \tilde{u}_\theta  +\frac{1}{B_{\phi,0}} \partial_r\left(\int^\theta \left [r \tilde{u}_r {B}_{\phi,0}\right]d\theta\right)
\label{Ohmslaw31}
\end{eqnarray}
Combined with equation (\ref{radfloworigin1}) results in an effective macroscopic toroidal projection of Ohm's law as \cite{krebsfluxp,Jorekhaoweifp1}:
\begin{eqnarray}
\fl -\frac{1}{R}\partial_t \psi_0 + \left(\langle\tilde{u}_r \tilde{B}_\theta\rangle +  \langle \tilde{B}_r \frac{1}{B_{\phi,0}} \partial_r\left(\int^\theta \left [r \tilde{u}_r {B}_{\phi,0}\right]d\theta\right) \rangle \right)= \eta J_{0,\phi} 
\label{macrotorohm11}
\end{eqnarray}
where $\psi_0 = R A_{\phi,0}$.
From the simulation data shown later on, this agrees almost perfectly for the later saturated phase. 

If we now consider the toroidal projection of equilibrium Ohm's law, i.e. the first of (\ref{jorekmhdeqequil1}), the dynamo is driven by the following correlations: $\langle \urt \Bpolt \rangle_\phi; -\langle \upolt \Brhot\rangle_\phi; \ur0\Bpol0$.
We need to compute these from the fundamental equations.
For the momentum equations, the equilibrium poloidal and toroidal flow generation is given by the Maxwell stress $\langle\mathbf{\tilde{J}}\times\mathbf{\tilde{B}}\rangle$ projected in the poloidal and toroidal directions. We will show that indeed a substantial flow is generated on top of the perturbed flow.
Finally, for the pressure evolution, third of equations (\ref{jorekmhdeqequil1}), we need the correlations $-\langle\mathbf{\tilde{u}}\cdot\nabla \tilde{P}\rangle_\phi; -\mathbf{u_0}\cdot\nabla P_0; -\frac{5}{3} P_0 \nabla\cdot \mathbf{u_0}$. The latter two require only $\ur0$, whereas the first perturbation correlation requires the knowledge of the flow and pressure perturbations.
\subsection{Global mode energy balance}
The perturbations system (\ref{jorekmhdeqpert1}) describes both the linear and the non--linear dynamics. We are interested here in understanding the energy balance of the MHD mode in both the linear and in the saturated phase.
If one multiplies the momentum equation with $\mathbf{\tilde{u}}\cdot$ and takes the flux--surface--average $\langle\rangle_{f.s.a.}$ one obtains the local evolution equation for the kinetic energy $E_K=0.5\rho_M \langle\tilde{u}^2\rangle_{f.s.a.}$ and for the magnetic energy $E_B=0.5\langle\tilde{B}^2\rangle_{f.s.a.}$ (vacuum permeability $\mu_0$ does not appear due to the chosen normalization):
\begin{eqnarray}
\fl \partial_t (E_K + E_B) = -\langle\mathbf{\tilde{u}}\cdot\nabla\tilde{P}\rangle_{f.s.a.} + \partial_V\langle (\partial_t \mathbf{\tilde{A}}+\nabla\tilde\Phi)\times\mathbf{\tilde{B}}\cdot\nabla V \rangle_{f.s.a.}-\langle \mathbf{J_0}\cdot(\mathbf{\tilde{u}}\times\mathbf{\tilde{B}}) \rangle_{f.s.a.} \nonumber \\
+ (t.c.t.)
\label{kinmagenergypertevol1}
\end{eqnarray}
where $(t.c.t.)$ means "triple correlation terms". The terms arising from the equilibrium flows have been neglected. The pressure term can also be rewritten in conservative form by discarding compressibility:
\begin{eqnarray}
\fl \partial_t (E_K + E_B) = -\partial_V\left[\langle \left(\tilde{P} \mathbf{\tilde{u}} - (\partial_t \mathbf{\tilde{A}}+\nabla\tilde\Phi)\times\mathbf{\tilde{B}}\right)\cdot\nabla V \rangle_{f.s.a.}\right]-\langle \mathbf{J_0}\cdot(\mathbf{\tilde{u}}\times\mathbf{\tilde{B}}) \rangle_{f.s.a.} \nonumber \\
+ (t.c.t.)
\label{kinmagenergypertevol11}
\end{eqnarray}
The pressure perturbation equation can instead be multiplied by $\tilde{P}$ to get ($E_P=0.5 \langle\tilde{P}^2\rangle_{f.s.a.}$):
\begin{eqnarray}
 \partial_t E_P = -\langle\tilde{P}\mathbf{\tilde{u}}\cdot\nabla P_0 \rangle_{f.s.a.} + (t.c.t.)
\label{presenergyperteq1}
\end{eqnarray}
Terms arising from transport/sources are also neglected.

To drive an unstable mode with positive energy growth, in the presence of peaked pressure profile ($-\partial_r P_0 > 0$), from (\ref{presenergyperteq1}) one must have $\langle\tilde{P}\mathbf{\tilde{u}}\cdot\nabla V\rangle_{f.s.a.} > 0$, that is, an outward flow of energy. This is very much analogous to the entropy principle in gyrokinetics \cite{gksugamaenthropy,garbetentropypaper,krommesentropygyrok}. Since $\tilde{P}$ must also follow the radial flow as $\tilde{P}\sim\urt (-\partial_r P_0)$, this means that in the kinetic+magnetic energy evolution equation (\ref{kinmagenergypertevol1}), the first term on the right--hand--side is bounded to be {\it negative}. 
But again, since this is a $\it growing$ mode, the right--hand--side must be positive as a whole, at each radius. 
Because of the conservative nature of the first two terms on the right--hand--side in equation (\ref{kinmagenergypertevol11}), the third and last term $-\langle \mathbf{J_0}\cdot(\mathbf{\tilde{u}}\times\mathbf{\tilde{B}}) \rangle_{f.s.a.}$ is identified as the {\it drive} of the mode. Locally, this term can be negative or positive. Close to the plasma center, away from the surface where the mode is excited, this term is negative and acts as a sink of energy (visible in the simulations data later on), which can be seen by the dominance of the equilibrium toroidal current term multiplying $\urt \Bpolt$, see section 3.1.1. However moving closer to the outer regions this term reverses sign and becomes positive, thus driving the mode. The main reason is the appearance of the equilibrium pressure gradient in the perpendicular component of the equilibrium current density and the field curvature from the parallel field fluctuations \cite{brunettimhdbasics,freidbergmhdbook1}, summarized here for the linear dynamics and without terms that are conservative:
\begin{eqnarray}
\fl -\langle \mathbf{J_0}\cdot(\mathbf{\tilde{u}}\times\mathbf{\tilde{B}}) \rangle_{f.s.a.} =\frac{1}{\gamma} \langle P_{\vert\vert} + P_{\perp} \rangle 
\nonumber \\
\fl P_{\vert\vert} = -\gamma\frac{J_{0,\vert\vert}}{B_0}\mathbf{B_0}\cdot(\tilde{\mathbf{u}}_{\perp}\times\tilde{\mathbf{B}}_{\perp}) \nonumber \\
\fl P_{\perp} = \frac{(\urt \partial_r P_0)^2}{B_0^2}+2 (\urt\partial_r P_0) \tilde{\mathbf{u}}_{\perp}\cdot\mathbf{\mathcal{K}} - (\urt\partial_r P_0) \nabla\cdot\tilde{\mathbf{u}}_{\perp}
\label{j0uxbtermsummary1}
\end{eqnarray} 
with $\mathcal{K}=(\mathbf{b}_0\cdot\nabla) \mathbf{b}_0$ is the magnetic field curvature vector. The term $P_{\vert\vert}$ is the parallel current drive, whereas the term $P_{\perp}$ is the pressure drive. Note that the quadratic term (first on the right--hand--side of the formula of $P_{\perp}$) cancels out with a term from the parallel magnetic energy time derivative $\gamma \tilde{B}_{\vert \vert}^2$. The same happens for the third term with the divergence of the perpendicular perturbed flow. As such, the actual driving term is $2 (\urt\partial_r P_0) \tilde{\mathbf{u}}_{\perp}\cdot\mathbf{\mathcal{K}}$, which is positive for typical pressure gradients embedded in a tokamak magnetic field with negative curvature on the low--field--side. In the center of the plasma the pressure gradient tends to zero, and there the parallel current term provides a stabilizing negative contribution.

Since the energy flux term is also a sink in the plasma center, the growing mode is sustained by the second term, the radial exchange term $ \partial_V\langle \partial_t \mathbf{\tilde{A}}\times\mathbf{\tilde{B}}\cdot\nabla V \rangle_{f.s.a.}$. In the linear phase, this has to be positive in the core and negative outside, since the volume integral of this term vanishes by assuming vanishing perturbations outside of, e.g. the $q=1$ surface or where the magnetic shear starts to increase strongly. The heat flow term discussed above conversely is negative in the core and positive outside (conservative due to incompressibility), thus acting as a redistribution of energy. Summarizing, the driving term $-\langle \mathbf{J_0}\cdot(\mathbf{\tilde{u}}\times\mathbf{\tilde{B}}) \rangle_{f.s.a.}$ is negative in the center of the plasma and becomes positive more outside where the pressure gradient drive sits (the link is provided by the eigenvalue equation for the interchange mode and its resulting pressure--driven growth rate). 

Mode saturation occurs when the flow perturbation decreases dramatically to make the term $\int dV (-\langle \mathbf{J_0}\cdot(\mathbf{\tilde{u}}\times\mathbf{\tilde{B}}) \rangle_{f.s.a.}) \rightarrow 0$.  But what causes this flow damping? In \cite{Jorekhaoweifp1} it is argued that the flow viscosity $-\mu\nabla^2\mathbf{\tilde{u}}$ leads to this damping. Indeed, on a long time scale, viscosity, resistivity and background energy transport all lead to damping of oscillations \cite{zhangmhdviscossawtfluxp}. However, on a shorter time scale, flux--pumping lives in the "collisionless" limit \cite{Jorekhaoweifp2}.  On this time scale, saturation is reached via two possible ways, either in sequence or concomitant. On one hand, when the pressure profile is rapidly modified so that the drive is reduced to lead the mode into the  stable region. On the other hand, if the non--linear coupling term $\widetilde{\mathbf{\tilde{J}}\times \mathbf{\tilde{B}}}$ is able to counteract the linear drive, saturation is also achieved. This latter mechanism involves moving energy to the higher toroidal harmonics. These two mechansisms are very much different than the saturation mechanisms of Larmor--scale turbulence, where the modification to the background profile is minimal on the time scale of evolution of the turbulent fields. The main difference being the amplitude of induced transport, which for MHD has as relevant scale the Alfv\'en speed, whereas for Larmor--scale turbulence is diamagnetic speed. For the latter phenomenology saturation is usually produced by the non--linear term and energy cascade to localized zonal flows \cite{zonalflowstabitgrogers1}. 
As mentioned already, the triple correlation term, e.g. non-linear coupling, can also lead to mode saturation as shown in \cite{rosenbluthnonlinmhdpaper1}.

\begin{figure}
\begin{center}
\includegraphics[width=90mm]{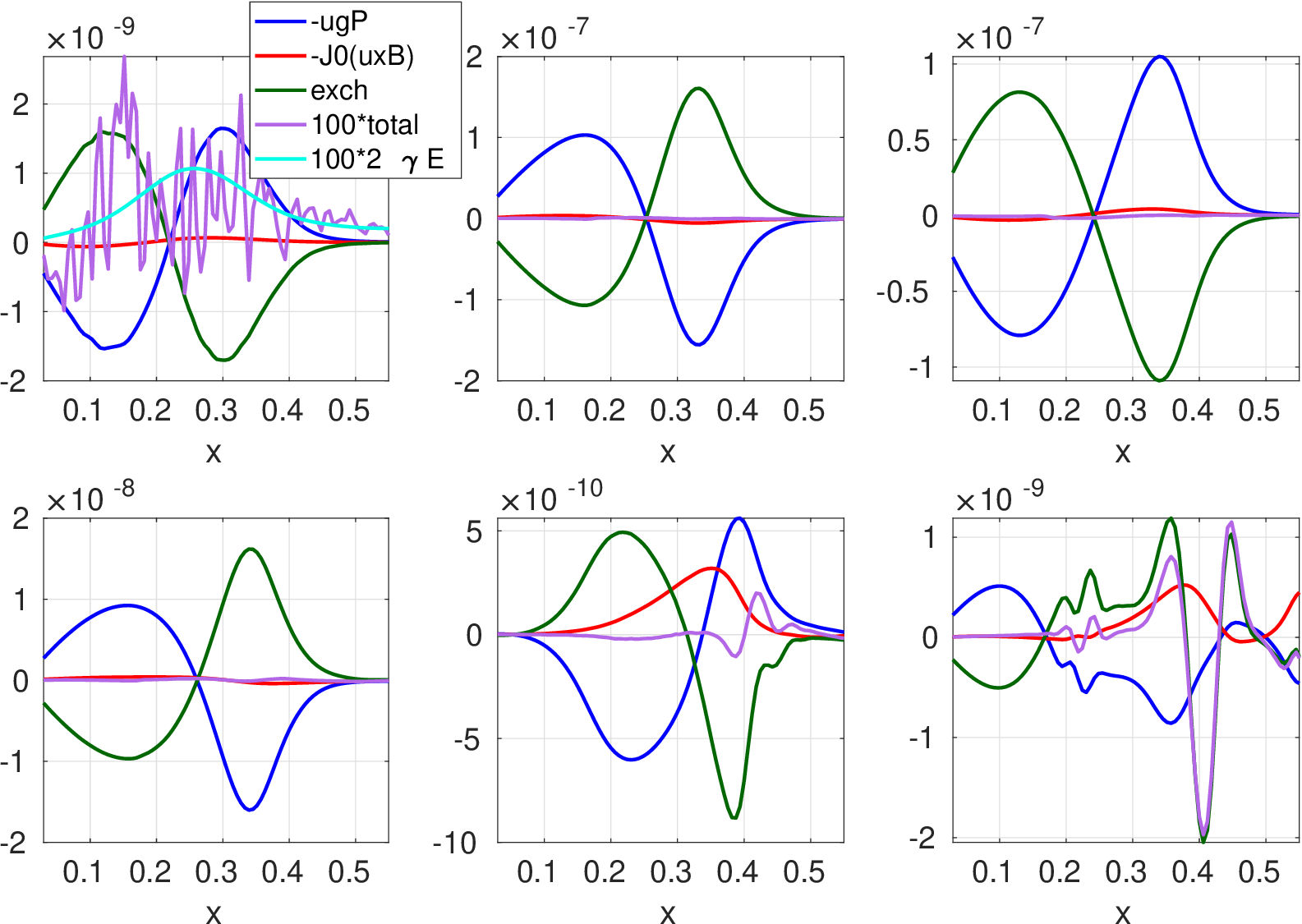}
\caption{\small Balance of the right--hand--side of equation (\ref{kinmagenergypertevol1}). The "total" curve is amplified by x100 in the first sub--plot on the left for sake of visibility. From top--left plot to bottom--right plot going clockwise: linear phase, early saturation/oscillatory phase (2nd, 3rd, 4th plot), saturated phase after 0.01 s (5th plot), late saturated phase (0.06 s, last plot).}
\label{figureenergybalancelinnonlin1}
\end{center}
\end{figure}

Another observation that can be made on equation (\ref{kinmagenergypertevol1}) is that at saturation, when the left--hand--side becomes near zero, on the right--hand-side the exchange term also becomes much smaller. The last dynamo power term drops to more negative due to the poloidal current $J_{\theta,0}$ becoming more negative at lower pressure gradient (from $\partial_r P_0 = J_{\theta,0}\Bphi0-J_{\phi,0}\Bpol0$, and the equilibrium toroidal field is negative $\Bphi0<0$ in this case). As such, the first term on the right--hand--side $-\langle\mathbf{\tilde{u}}\cdot\nabla\tilde{P}\rangle_{f.s.a.}$ is expected to become small and positive now, e.g. an influx of heat into the core. 
Before moving to a reduced model of mode saturation, we plot the terms in equation (\ref{kinmagenergypertevol1}) from a JOREK simulation to clearly show visually the discussion just presented. In figure \ref{figureenergybalancelinnonlin1} this balance is presented. It can be clearly seen that the pressure and the dynamo exchange term balance almost exactly, but with a small positive remnant which drives the mode (the growth rate is much smaller than the Alfv\'en speed). In the saturated phase (second and third sub--plots), the total is very much oscillating around zero although there are still large opposite contributions from the pressure drive and the dynamo drive.

We build now a simple 1D in radius and time--dependent predator--prey like model that describes this dynamics and check if the resulting time--scale is compatible with the simulation results. The model is presented here and the comparison to JOREK will be given in Section 4.
\subsection{Simplified 0D model of the mode saturation}
The predator--prey like model that we build, inspired by the simple edge--localized--modes model presented in \cite{mhdprepreyoldmodel1costantin,Hzohmbookmhd}, is expanded to the following 4--fields system:
\begin{eqnarray}
 \partial_t P = H - \chi P - \hat{\chi}u p \nonumber \\
 \partial_t p = P u  \nonumber \\
 \partial_t B = b u \nonumber \\
 \partial_t u = -\mu u +B (a P - b) 
\label{predpreymodel1}
\end{eqnarray}
where the dynamical variables are the equilibrium pressure $P$, the perturbed 
pressure $p$, the radial flow $u$ and the perturbed magnetic field $B$. The 
term $H$ is a constant heating source, $\chi$ is the background transport given by e.g. turbulence + neoclassical, $\hat{\chi}$ is a constant 
coefficient that provides MHD transport of heating, $b$ is the coupling 
constant between flow and magnetic field (field line bending), $\mu$ the flow 
damping rate via viscosity, $a$ a constant coupling the pressure drive to the 
kinetic flow equation. 
The normalization of the quantities in this set of simplified equations are the same as used in the JOREK code. 
In model (\ref{predpreymodel1}) the non--linear 
terms in the $p$, $B$ and $u$ equations are not included since, as discussed 
previously, the main saturation mechanism is thought to be the modification of the equilibrium pressure profile $P$ towards marginal stability. 
Note that in a recent work it has been found an example of simulation in which the nonlinear term produces saturation before any visible modification to the equlibrium profiles \cite{vmecmegasaturatedkinkinterchangecomp1}, thus both situations can indeed take place. The discriminant is probably the strength of the perturbed flow compared to the strength of the perturbed current and field. If such term maybe added, it could appear in the third of equations (\ref{predpreymodel1}) as $\partial_t u = ... - \alpha B^3$ with $\alpha$ a numerical constant \cite{nonlinearinterchange1walbr}.

If one compares this simple model for the fluctuations to the actual MHD equations system (\ref{jorekmhdeqpert1}), can recognize a few term, however in the momentum equation some modification has been made. The sound--wave term $-\nabla\tilde{P}$ and the field energy term $\mathbf{\tilde{J}}\times\mathbf{B_0}$ are not considered since they are much larger than the actual driving term $\mathbf{J_0}\times\mathbf{\tilde{B}}$, and they balance each other out, moreover they end up being conservative terms, as discussed already in subsection 2.1. The driving term $\mathbf{J_0}\times\mathbf{\tilde{B}}$ is represented as $B(aP-b)$, where the $b$ stabilizing term comes from the field line bending and the field energy terms in equations (\ref{kinmagenergypertevol11}) and (\ref{j0uxbtermsummary1}). This also acts as the coupling between the momentum equation and Faraday's law. The radial fluctuation force balance is $\partial_r \tilde{P} \approx \Jpolt \Bphi0$. This is guarantueed since $\partial_r \tilde{P}\approx \urt \partial_r \lambda$ and $\Jpolt \approx -\partial_r\Bphit$ and $\Bphit \approx -\urt \lambda/\Bphi0$, see \cite{brunettimhdbasics}, equation (7.35), first term on the right--hand--side. The other terms contribute to the drive.

The linear instability is obtained by starting from $P_0 = H/\chi \ge b/a$, with resulting growth rate $\gamma=\sqrt{b(a P -b)}$, whereas we also define the  growth rate provided by the initial equilibrium pressure as $\gamma_0=\sqrt{b(a P_0 -b)}$. The linear phase ends when the energy flow term $- \hat{\chi}u p$ starts to reduce the pressure $P$. Then the saturation of the MHD mode starts. We now study this dynamics.
First, we notice that the 2nd and 3rd equation of the system (\ref{predpreymodel1}) are identical, both driven by the flow $u$, with the different rate constant $b$ for the magnetic field and $P$ for the pressure. Since we anticipate that it is observed that the pressure $P$ does not change dramatically during the evolution, we can approximately reduce the system by one field in doing:
\begin{eqnarray}
 p \approx \frac{P}{b} B  \nonumber \\
 \partial_t B = b u \nonumber \\
 \partial_t u = -\mu u +B (a P - b) \nonumber \\ 
 \partial_t P \approx H - \chi P - \hat{\chi}u B
\label{predpreymodel111}
\end{eqnarray}
where $\hat{\chi}$ has absorbed the $P/b$ factor.
The approximation used here is valid as long as both the equilibrium pressure nor the equilibrium magnetic field are not strongly modified by the mode along the linear and early saturation phase. This means that the pressure profiles sits already quite close to threshold, and the long-time impact of the mode on the evolution of the safety factor and the pressure has not yet kicked in. 

We now define $\xi = B/b, k = ab$, and upon looking at the system (\ref{predpreymodel111}), one can derive a single equation for $\xi$ as:
\begin{eqnarray}
\partial^2_t \xi = \gamma^2 \xi - \mu \partial_t \xi
\label{nonlinoscillat1}
\end{eqnarray}
Then we define the pressure variation $\delta = P-P_0$ and the equilibrium power source $H=\chi P_0$ to manipulate the pressure equation into:
\begin{eqnarray}
\partial_t (\delta + \frac{b\chihat}{2}\xi^2) = -\chi \delta
\label{nonlinoscillat2}
\end{eqnarray}
which is formally solved as $\delta = -\frac{b\chihat}{2}\xi^2 + C(t)$, with $C(t)=-\chi\int \delta dt $. 
We can thus obtain the momentum equation rewritten as:
\begin{eqnarray}
\partial^2_t \xi = \gamma_0^2 \xi -\frac{k b \chihat}{2}\xi^3 + k C(t)\xi - \mu \partial_t \xi
\label{nonlinoscillat3}
\end{eqnarray}
with $\gamma_0^2 = k(P_0-b/a) > 0$.
The term $C(t)$ is proportional to the background transport coefficient $\chi$. As such we now discuss two different cases.
\subsubsection{No transport}
If the transport of pressure on confinement time scale from $\chi$ would not count (as well as the power source), then $C=0$ and equation (\ref{nonlinoscillat3}) would acquire the character of a non-linear Duffing--like damped oscillator which maximal linear frequency is $\omega_{\rm max} = \sqrt{2}\gamma_0$, with damping provided by viscosity $\mu$. The neutral equilibrium points would be $\xi_{\rm eq}=0$ and $\xi_{\rm eq}=\sqrt{\frac{2}{k b \chihat}}\gamma_0$. The non--linear oscillator frequency decreases from $\omega_{\rm max}$ to zero as the oscillation amplitude increases due to the non--linear term slowing down the oscillation as it approaches the null neutral point away from the finite one.

However the non--linear frequency at zero transport is calculated to be below what is observed from the JOREK simulations, figure \ref{figure1}, where it can be clearly seen that the estimated oscillation frequency is larger than $\omega_{\rm max}$, closer to $\approx 2\gamma_0$.
\subsubsection{With transport}
Now adding back the transport part $H-\chi P$, we solve analytically equation (\ref{nonlinoscillat2}) as $\delta = -\frac{b\chihat}{2}\xi^2+\frac{b\chi \chihat}{2}\int_{-\infty}^t e^{-\chi(t-s)}\xi^2(s) ds$. 
The new neutral equilbrium point that is not zero is thus formally: $\xi_{\rm eq}=\sqrt{\frac{2}{k b\chihat}}\left(\gamma_0^2+\frac{k b \chi \chihat}{2}\int_{-\infty}^t e^{-\chi(t-s)}\xi^2(s) ds\right)^{1/2}$. Note that the new term in the parenthesis is positive definite. The expression is implicit since $\xi$ appears both as solution but also inside the integration kernel.
Linearizing again equation (\ref{nonlinoscillat3}) around the new neutral point gives a frequency:
\begin{eqnarray}
\omega^2=2\gamma_0^2+b k \chi\chihat\int_{-\infty}^t e^{-\chi(t-s)}\xi^2(s) ds
\label{nonlinoscillat4}
\end{eqnarray}
Note that since the transport time scale set by $\chi$ is much longer than either the linear growth phase scale and the oscillatory phase, we approximate $\chi \approx 0$ in the exponent inside the integral.
An estimate can then be obtained by considering that the memory effect requires taking into account the whole linear phase, plus to compute the frequency over 1 period, we need to integrate again up to half-period, in this way:
\begin{eqnarray}
\fl \int_{-\infty}^t \xi^2(s) ds \approx \int_{-\infty}^{0} \xi^2(s) ds + \int_0^{T/2}\xi(s)^2 ds\approx \frac{2\gamma_0}{k b \chihat}+\frac{\pi}{\omega}\frac{2 \gamma_0^2}{k b \chihat}.
\label{nonlinoscillat41}
\end{eqnarray}
where we have used, for the linear phase, the exponential form for the eigenvector $\xi \sim e^{\gamma_0 t}$ and using for the value at $s=0$ the maximal one $\xi^2=4\gamma_0^2/(k  b \chihat)$. For the first half period we have instead assumed a simple linear decrease from $\xi^2=4\gamma_0^2/(k  b \chihat) \rightarrow \gamma_0^2/(k  b \chihat)$, which mean value is $2\gamma_0^2/(k  b \chihat)$. The half period is simply $\pi/\omega$. Finally:
\begin{eqnarray}
\omega = \gamma_0\sqrt{2}\sqrt{1+\frac{\chi}{\gamma_0}\left(1+\frac{\gamma_0\pi}{\omega}\right)} 
\label{nonlinoscillat5}
\end{eqnarray}
Lastly, since $\chi$ adds a small correction, by recurrence at the lowest order we approximate $\omega \approx \sqrt{2}\gamma_0\sqrt{1+3.22\frac{\chi}{\gamma_0}}$.
This frequency is larger than $\sqrt{2}\gamma_0$ by a sizeable offset. 

\subsubsection{Comparison with the JOREK simulations}
\begin{figure}
\begin{center}
\includegraphics[width=70mm]{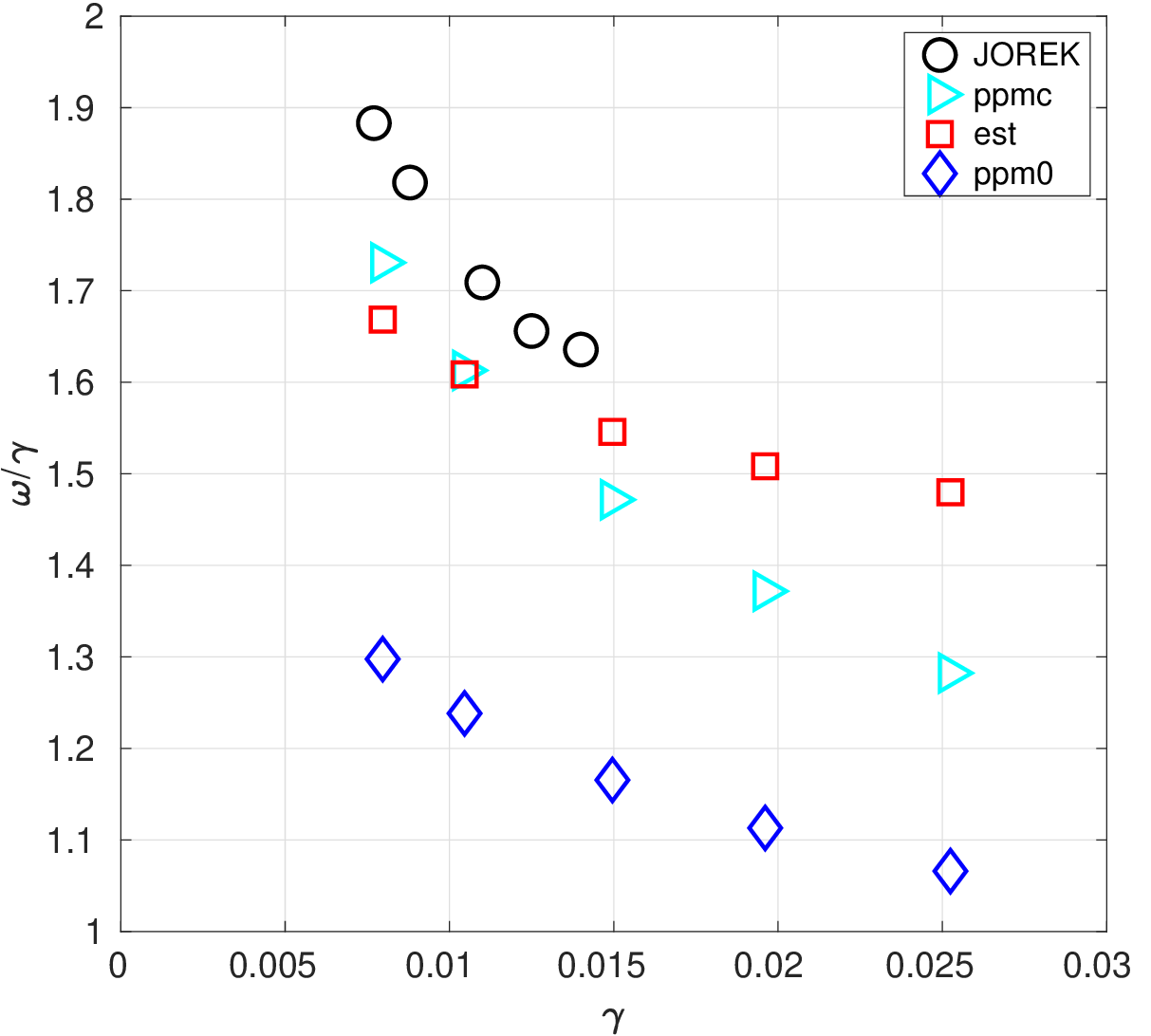}
\caption{\small Comparison of the oscillation frequency at saturation between a JOREK scan ("JOREK", black circles), the solution of the predator--prey system for a similar scan ("ppmc", cyan triangles) and the analytical estimate from equation (\ref{nonlinoscillat5}) ("est", red squares). In addition a scan with the predator--prey model where the transport coefficient $\chi=0$ is also shown ("ppm0", blue diamonds).}
\label{figure1}
\end{center}
\end{figure}

In \cite{Jorekhaoweifp1} for a simulation case it is quoted a growth rate of $\gamma \approx 19460 s^{-1}$ and an oscillation frequency of $f \approx 5.8$ kHz, which ratio is $2\pi f/\gamma\approx 1.9$ is higher than $\sqrt{2}$ and closer to the present estimate.
To this purpose, a transport coefficient scan has been carried out, and the analytical considerations presented before explain the obtained JOREK results as shown in figure \ref{figure1}, where the observed frequency is closer to $2\gamma_0$ rather than $\sqrt{2}\gamma_0$. The ratio becomes indeed $\approx \sqrt{2}\gamma_0$ for the highest growth rates, where the effect of transport is less relevant. The physics is that transport will tend to raise the equilibrium pressure as soon as it drops, due to the presence of the power source $H$ and the constancy of $\chi$ (although in reality this diffusivity would also drop due to the lowered gradient). As such the MHD mode is now driven again at a stronger rate than if the background transport was not present at all. 
In figure \ref{figure1} we compare the growth rate variation for a scan performed in JOREK varying the background transport coefficient $\chi$, the oscillation frequency measured in the simulation versus the analytical estimate (\ref{nonlinoscillat5}). 
In the figure we also show the same calculation but setting the transport coefficient $\chi=0$. In that the case the frequency is lower as expected. The agremeent between the solution of the predator--prey model and the simple analytical estimate are closer for the lowest growth rates where the non--linear effects ($\sim \xi^3$) are less important.

Note that how the system behaves, it always leads to saturation as soon as the pressure drops below the critical value. To get a sawtooth--like behavior one would need an additional mechanism that breaks the surfaces once they are too compressed by the radial displacement. This mechanism is recognized to be magnetic reconnection, which is not part of this model. Artificially, by playing with the parameters it is in principle possible to get a major crash of the pressure, however physically this is not what happens in reality. Also in the case of edge instabilities, the major crash is terminated by a removal of the edge layer, via magnetic reconnection and formation of individual 
filaments \cite{ELMshzohm}.

If saturation would be due to non--linear coupling \cite{nonlinearinterchange1walbr} setting in before the equilibrium pressure changes, then the same kind of eigenfrequency analysis can be carried out with the addition of the term $-\alpha B^3$ as mentioned previously, to the momentum equation. In this case it can be proven that the oscillation frequency after the linear phase would be $f \approx \sqrt{2}\gamma$, since it does not rely on transport evolution, instead of $\approx 2\gamma$. Indeed this seems to be the case in the simulations reported in \cite{vmecmegasaturatedkinkinterchangecomp1}, where saturation is reached before the equilibrium pressure or safety factor begin to evolve. It would be interesting to check if this happens experimentally too.

\begin{figure}
\begin{center}
\includegraphics[width=70mm]{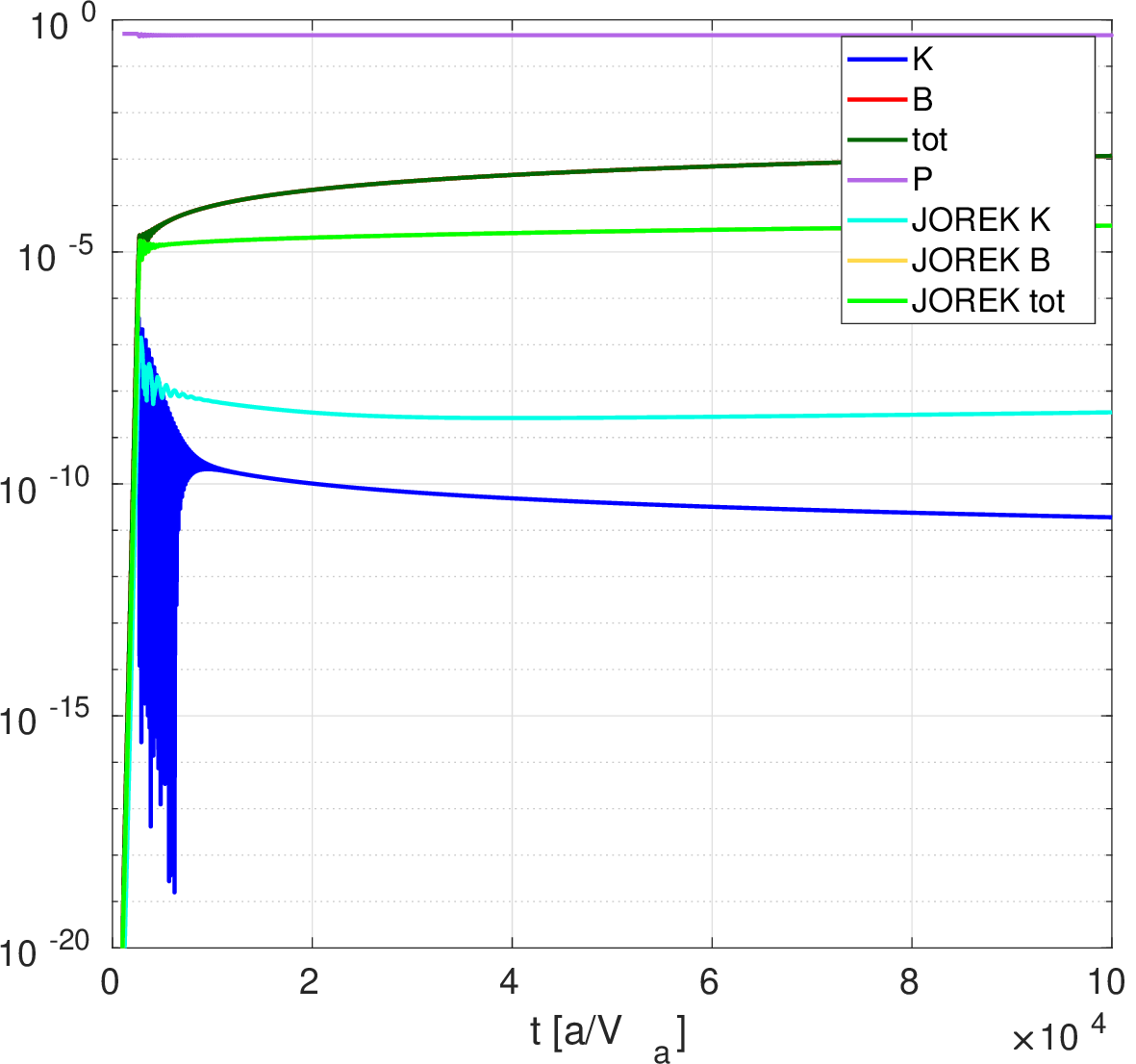}
\includegraphics[width=70mm]{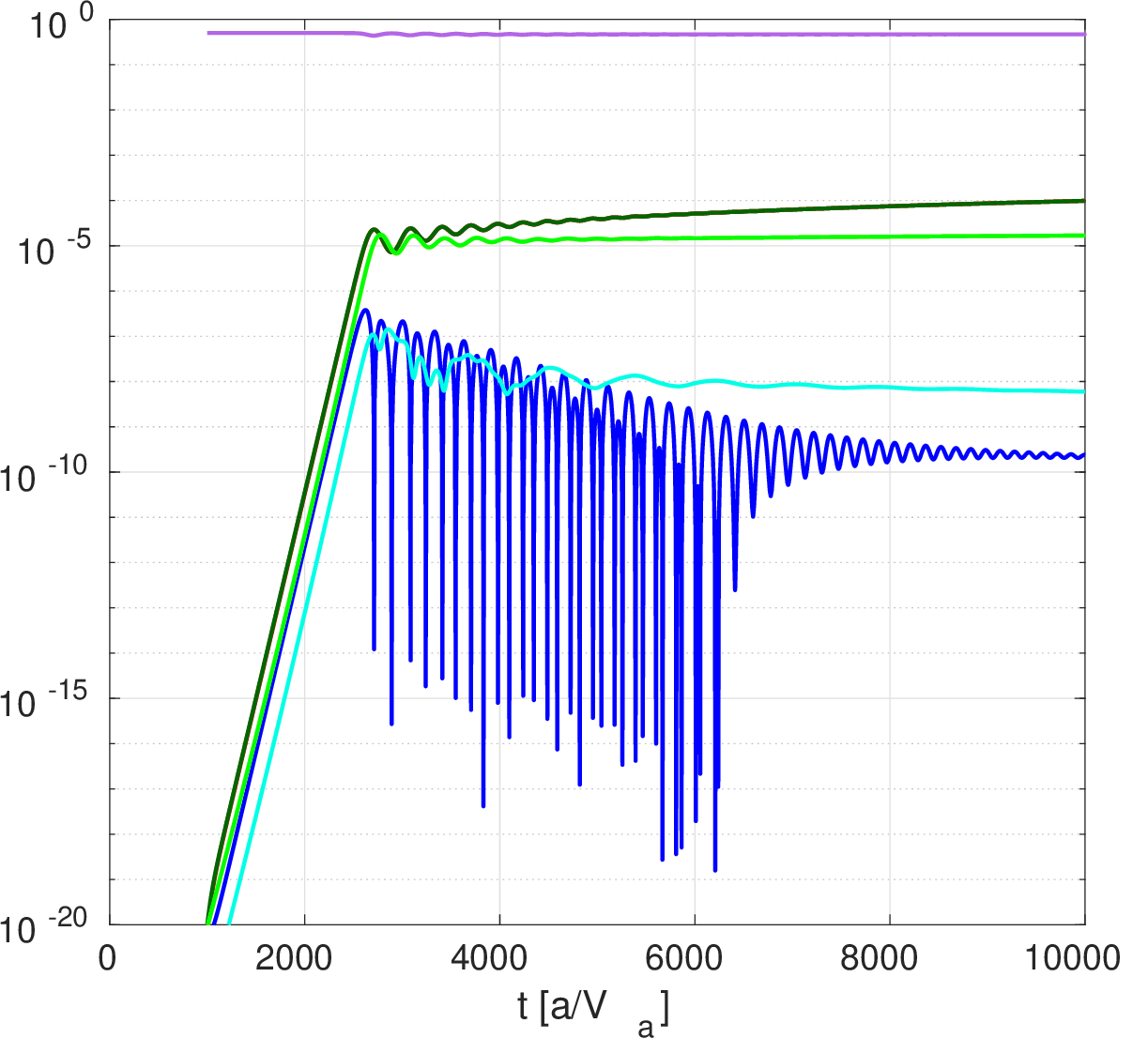}
\caption{\small Kinetic energy $K$, magnetic field energy $E_B$, and sum of the two "tot" from simple model and from JOREK ("JOREK K", "JOREK $E_B$", "JOREK tot"). The plot on the right is a zoom to show in details the oscillatory phase after the linear phase.}
\label{figure2}
\end{center}
\end{figure}

\begin{figure}
\begin{center}
\includegraphics[width=70mm]{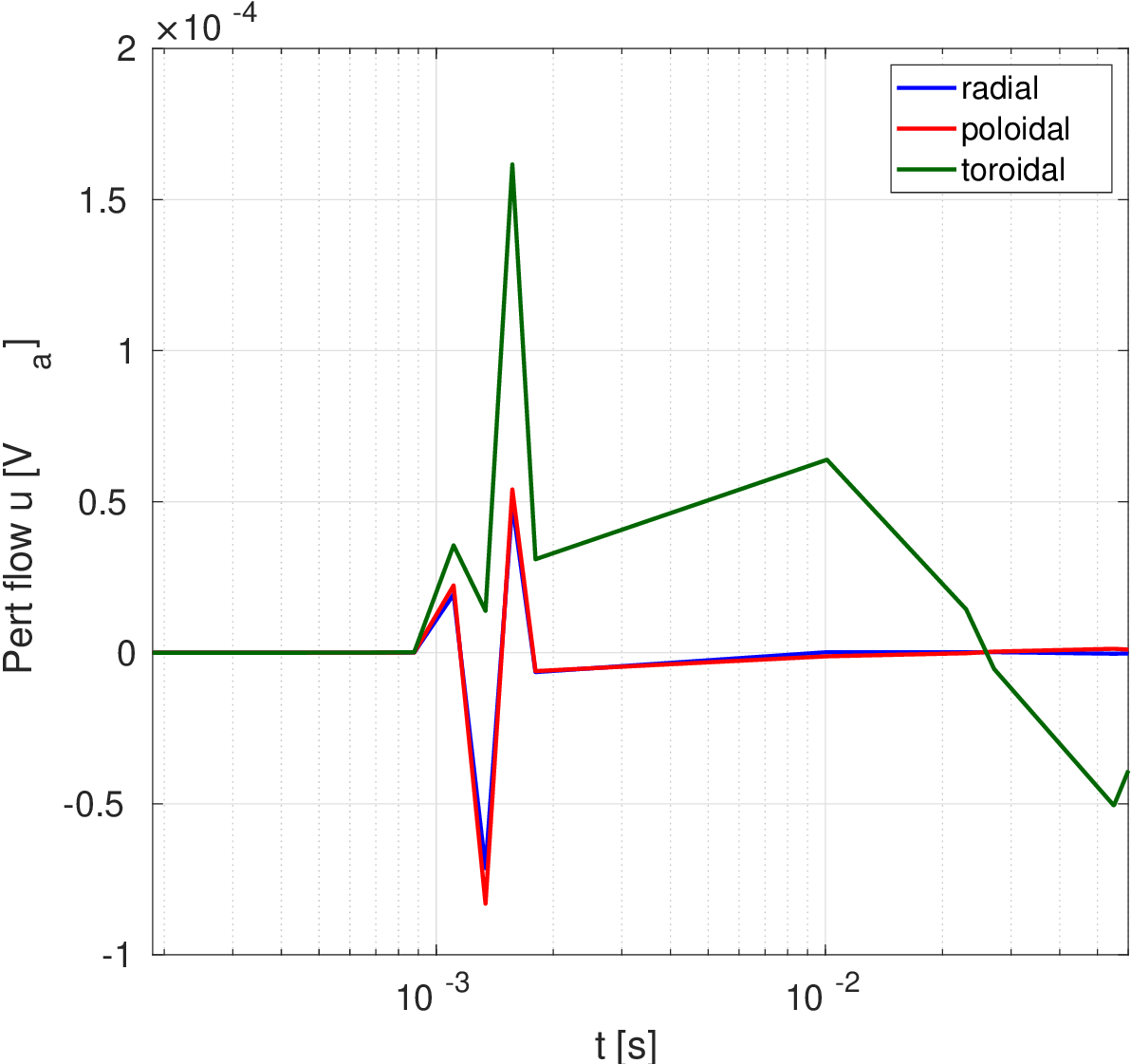}
\caption{\small $n \neq 1$ flow components radial $\urt$ (blue), poloidal  
$\upolt$ (red) and toroidal $\uphit$ at $x \approx 0.085$, $\theta=0,\phi=0$, 
with $x$ normalized poloidal 
flux coordinate, as a function of time.}
\label{figure3}
\end{center}
\end{figure}

In figure \ref{figure2} we compare the model (\ref{predpreymodel111}) to an actual JOREK run. The model parameters used to match the simulation are: $a=0.086, b=0.04, \chi=0.001, \hat{\chi}=200, H=0.0005$. The resulting initial pressure is $P_0=0.5$ and the threshold is $b/a = 0.4651$. The resulting growth rate is $\gamma=0.011$. Note that time is normalized to Alfv\'en time $V_a/a$ with $a$ the minor radius and $V_a$ the Alfv\'en speed.
The kinetic energy from the model is defined as $K=0.5 u^2$, the magnetic energy $E_B$ as $E_B=0.5 B^2$.
Since the simple model (\ref{predpreymodel1}) has only 1 flow and 1 magnetic field components, and the steady state solution implies a decaying flow, it explains the result shown in figure \ref{figure2} where the kinetic energy $K$ decays to much lower values than the JOREK simulation. In the code, as in the actual MHD equations, the flow has 3 components, and while the radial component becomes indeed very weak in the saturated phase, both the poloidal and toroidal perturbed flows remain substantial, linked by the non-linear phase relation $\uphit\Bpol0\approx\upolt\Bphi0$. As such the simulation kinetic energy is sustained by the non--radial flow components.
This can be seen in figure \ref{figure3}, 
where the radial component follows well the poloidal component until after the oscillatory phase, after which $\urt \rightarrow 0$ whereas the poloidal and toroidal flow components now are correlated.

Note also that according to the full model for the perturbations, equations (\ref{jorekmhdeqpert1}), only non--ideal effects can allow a finite perturbed flow 
pattern at fixed perturbed fields, finite perturbed flow that can explain the 
magnitude of the dynamo term observed in \cite{Jorekhaoweifp1}. The physics here is that there are two very different flows: one that drags the magnetic field and one that drags the plasma particles. Both coincide in the standard "frozen--in--law" case of ideal MHD. However, in the presence of non--ideal effects, the two are different, and the plasma can move at a different speed of the field lines. The terms with the 
equilibrium flow $\mathbf{u}_0$ do not count since they only produce a rotation of 
the mode (and the radial component is negligibly small). Resistive terms also are negligibly small and only count on equilibrium time scales. That all remains is the non--linear terms, for example $\tilde{u}\times\tilde{\mathbf{B}}$. Technically, the nonlinear terms effect on the fluctuation always appear as $\tilde{f}_1 \tilde{f}_2 - \langle \tilde{f}_1 \tilde{f}_2 \rangle_\phi$, i.e. subtracting the 0/0 component obviously. This however means that the non--ideal effects related to the non--linear coupling can only appear via e.g. (-1/-1) + (2/2) coupling producing a net effect on the 1/1 mode. As such, the persistence of flows in the long--time scale saturated phase is due to the coupling of different poloidal/toroidal harmonics and the complex conjugate of others. Moreover, the non--linear terms also allow the mode to saturate at a level which can be still slightly unstable, but which free energy is compensated by the energy redistribution in the higher harmonics rather than in driving the mode.  
\section{Linear to early saturation phase and fast dynamo}
In the previous Section we have discussed the linear and saturation phases of the mode at a global level. Regarding the $n \neq 1$ perturbations, it is clear that in the linear phase the analysis is simplified by the fact that every time derivative can simply be substituted by the mode complex frequency: $\partial_t \rightarrow -i \omega = \gamma - i \Omega$ with $\gamma$ the growth (or damping) rate and $\Omega$ the real rotation frequency, together with toroidal and poloidal derivatives being replaceable by a reduced set of dominant wavenumbers. 

In the long--time scale non--linear self--organized state this is not doable, however there are other methods that one can use. For now we limit ourselves to the linear analysis and provide some ideas for the non--linear phase, which will be addressed more in details in a future work. In any case, a discussion on the early saturation phase, where the "fast dynamo" is observed
\cite{Jorekhaoweifp1},  is provided.
\subsection{Dynamo term from radial flow perturbation}
We begin by considering the dynamo term arising from $\langle\urt\Bpolt\rangle_{f.s.a}$ (note that the flux--surface--average includes the toroidal average).  
\subsubsection{Linear phase}
We express $\gamma\Bpolt \approx \mathbf{B_0}\cdot\nabla\upolt+\urt\Bpol0/r-\urt\partial_r\Bpol0$ (incompressible flow assumed) to get:
\begin{eqnarray}
\gamma\langle \urt \Bpolt \rangle_\phi= \langle \urt \mathbf{B_0}\cdot\nabla\upolt \rangle_\phi - \langle \urt^2 (\partial_r \Bpol0-\Bpol0/r) \rangle_\phi
\label{urbtlinear1}
\end{eqnarray}
and using $\nabla\cdot\mathbf{u}\approx 0 \rightarrow \upolt\approx i/m \; \partial_r(r\urt)$ and $\mathbf{B_0}\cdot\nabla \rightarrow i (\Bphi0/R)(m/q + n)$ and correcting for complex conjucation we get (for $r \rightarrow 0$):
\begin{eqnarray}
\langle \urt \Bpolt \rangle_\phi= \frac{1}{\gamma}\langle \urt^2\left(-\frac{\Bphi0}{R}(1/q +n/m) + \frac{\Bpol0}{r} - \partial_r \Bpol0 \right)\rangle_\phi
\label{urbtlinear2}
\end{eqnarray}
In the convention in which $\Bphi0<0, \Bpol0>0 \rightarrow q < 0 $, and the mode has helicity $n=1,m=1$. If $\vert q\vert > 1$ then the $\Bphi0(m/q + n) < 0 $ and thus the parenthesis is positive ($\partial_r \Bpol0$ is small for flat $q$ profile). Then this term acts as an positive dynamo and pushes $\vert q\vert $ below unity.
Instead, if $q$ is monotonic and $\vert q \vert <1$, then the entire term in formula (\ref{urbtlinear2}) becomes negative and acts as a negative dynamo to sustain $\vert q \vert \approx 1$. The functional form of $\urt$ is determined at this stage by the solution of the mode eigenvalue equation.
\subsubsection{Fast dynamo}
We have concluded that when the safety factor $q$ (in absolute value) drops below unity, this term is negative, thus leading to a dynamo that keeps the safety factor from dropping even further below unity. 
At the end of the linear phase, the effective growth rate $\gamma$ appearing in formula (\ref{urbtlinear2}) drops to $\approx 0$.  
As a matter of fact, during the dynamics identified in Section 2.2, equations (\ref{predpreymodel111}-\ref{nonlinoscillat4}), e.g. the fast oscillatory phase, $\gamma$ continues to change sign, and similarly the radial flow $\urt$ oscillates positive and negative alike. However, the magnetic field oscillates weakly around a rather constant value, as discussed regarding equations (\ref{nonlinoscillat1}-\ref{nonlinoscillat4}). As such, the sign of $\langle \urt \Bpolt \rangle_\phi$ will strongly oscillate as well between negative and positive values, thus explaining the fast dynamo observed in the  numerical simulations. Similarly the pressure is affected in the same way by a rapidly oscillating term.
\subsubsection{Poloidal contribution}
The correlation $-\langle\upolt\Brhot \rangle_{f.s.a.}$ contributes to the total dynamo term as well. It is similar to $\langle\urt\Bpolt \rangle_{f.s.a.}$ in the linear phase, whereas it becomes much larger in the non--linear phase, but is mostly balanced by the equilibrium flow term discussed below.  
\subsection{Equilibrium radial flow from equation (\ref{radfloworigin1})}
We need to link the correlation $\langle\uphit\Brhot \rangle_\phi$ to the mode properties to understand how the equilibrium flow $\ur0$ is generated. 
\subsubsection{Linear phase}
Analogous to the dynamo term, we can employ a single $n/m$ mode structure and use the parallel projection of the momentum equation to find:
\begin{eqnarray}
\rho_M \Bphi0 \partial_t \uphit\approx - \mathbf{B_0}\cdot \nabla\tilde{P} - \Brhot\partial_r P_0
\label{uphieqbr1}
\end{eqnarray}
where we have neglected the poloidal flow contribution.
Assuming that the safety factor $q$ is flat and close to unity, we can neglect the parallel derivative and thus find:
\begin{eqnarray}
\rho_M \gamma \Bphi0 \uphit\Brhot \approx - \Brhot^2\partial_r P_0
\label{uphieqbr2}
\end{eqnarray}
Since the pressure gradient is negative, this means a positive $\Bphi0 \uphit\Brhot$, thus (since $\Bphi0<0$ in our convention), leades to a positive $\ur0$.
\subsection{Energy flow terms}
The equilbrium pressure evolution equation is:
\begin{eqnarray}
\partial_t P_0= -\langle \mathbf{\tilde{u}}\cdot \nabla\tilde{P}\rangle_{f.s.a.} -\frac{5}{3}P_0 \partial_V \langle\ur0\vert\nabla V\vert\rangle \rangle_{f.s.a.} + Tr + \mathcal{P}
\label{eqpresseq1}
\end{eqnarray}
where terms arising from Larmor--scale turbulence and neoclassical transport are not discussed but should be included as well. In the long time scale saturated phase of the quasi--interchange mode, we have established that the term $-\langle \mathbf{\tilde{u}}\cdot \nabla\tilde{P}\rangle_{f.s.a.} \rightarrow 0$ or slightly positive, as the self--organizing state is close to marginal stability or slightly stable, whereas $-\frac{5}{3}P_0 \partial_V \langle\ur0\vert\nabla V\vert\rangle \rangle_{f.s.a.}$ is robustly negative in the core and positive outside, thus providing a continous redistribution of energy from the inside to the outside. This is analogous to a sawtooth 
crash but softer.
\subsubsection{Linear phase}
A simple substitution gives
\begin{eqnarray}
\urt\partial_r \tilde{P} = \frac{1}{\gamma}\urt\partial_r (\urt \lambda)  =\frac{1}{\gamma}\left( \lambda\partial_r\frac{\urt^2}{2}  + \urt^2\partial_r\lambda\right) 
\label{enflowcorr1}
\end{eqnarray}
with $\lambda = -\partial_r P_0 > 0$ the equilibrium pressure gradient. Since the radial gradient of $\lambda$ is also positive, the right--hand side of formula (\ref{enflowcorr1}) is positive if the profile of $\urt$ is flat in the core where the mode is active. In the case of the internal kink it would be the case, whereas for the quasi--interchange is decreasing from the axis to the $q=1$. As such the first term on the right--hand--side is negative, but smaller than the second term, leaving a net positive energy flux from the core to the outside.
\subsubsection{Fast pressure oscillations}
As discussed in subsection 3.1.2, one can expect a similar behavior for the equilibrium pressure profile, being pushed up and down by a rapidly varying radial flow whereas the pressure fluctuation oscillates weakly around its saturated value. That is, the term $\langle \mathbf{\tilde{u}}\cdot \nabla\tilde{P}\rangle_{f.s.a.}$ oscillates strongly since $\urt \rightarrow \pm$, whereas $\tilde{P} \approx const$.
\section{Comparison to JOREK simulation results}
To assess the quality of the predictions of the various correlations discussed before, we compare to simulations carried out with the JOREK code,  very similar to what presented in \cite{Jorekhaoweifp1}. We have selected a typical case that produces flux pumping and displays all the features discussed above, i.e. the same case shown in figure \ref{figure2}. 
In Section 2 we have already compared a simple predator--prey type model against a JOREK simulation to display salient features of the saturation dynamics. Here we instead compare the actual estimates of Section 3 in the saturated phase.

\begin{figure}
\begin{center}
\includegraphics[width=70mm]{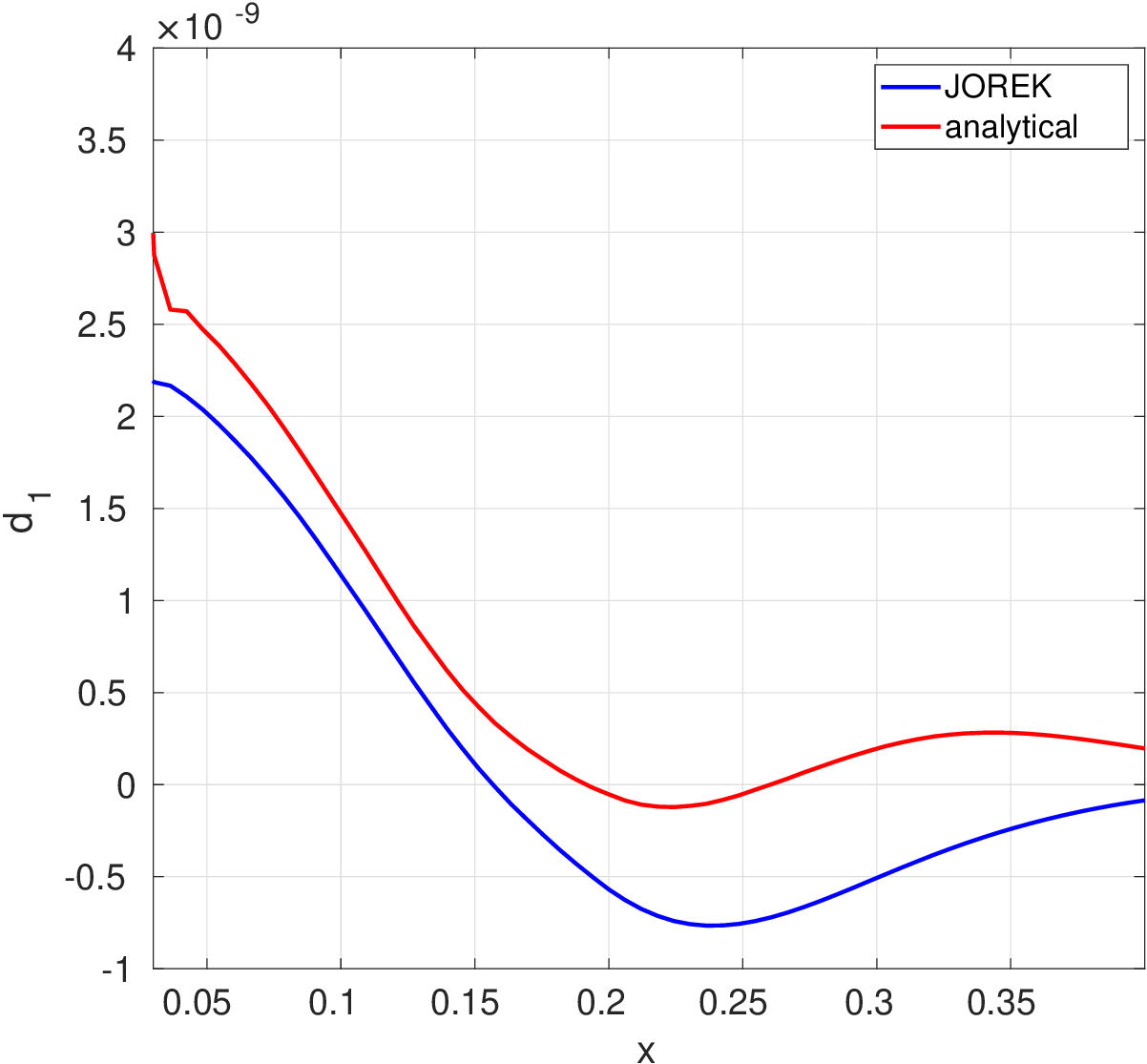}
\includegraphics[width=70mm]{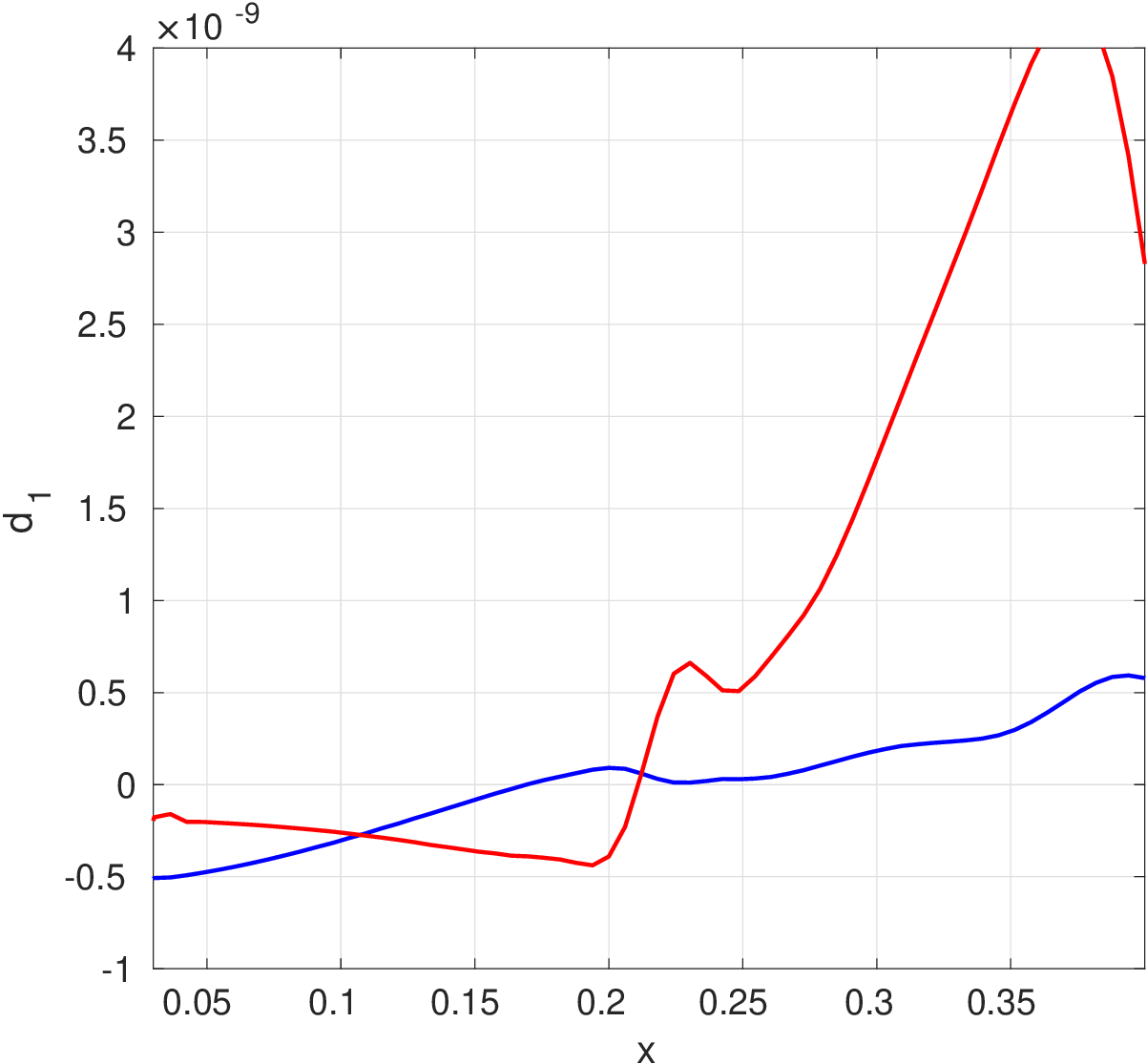}
\caption{\small Comparison of the dynamo term $d_1$ between analytical expression (\ref{urbtlinear2}) and JOREK result: (left) linear phase, (right) saturated phase.}
\label{figure4}
\end{center}
\end{figure}

In figure \ref{figure4} we first compare formula (\ref{urbtlinear2}) with the code result for the dynamo term $d_1=\langle \urt \Bpolt \rangle_\phi$.
It can be seen that the linear result agrees very well both quantitatively and qualitatively, whereas in the non--linear phase its only qualitatively in sign, whereas the shape is different. In magnitude for the saturated phase we had to reduce $\gamma \rightarrow 0.01 \gamma_{\rm linear}$. This is indicative of the quasi--saturated phase, however on a longer time scale one would expect a vanishing growth rate but a finite field perturbation. This is clear by taking equations (\ref{jorekmhdeqpert1}) and setting the time derivative to zero (and assuming no equilbrium flow):
\begin{eqnarray}
\fl 0=-\nabla\tilde{\Phi} + \widetilde{\mathbf{\tilde{u}}\times \mathbf{\tilde{B}}}+ \mathbf{\tilde{u}}\times \mathbf{B_0}  \nonumber \\
\fl 0 = -\nabla \tilde{P} + \widetilde{\mathbf{\tilde{J}}\times \mathbf{\tilde{B}}}+ \mathbf{J_0}\times \mathbf{\tilde{B}}+ \mathbf{\tilde{J}}\times \mathbf{B_0} + \mu \nabla^2 \mathbf{\tilde{u}} \nonumber \\
\fl 0 = -\mathbf{\tilde{u}}\cdot\nabla P_0-\widetilde{\mathbf{\tilde{u}}\cdot\nabla \tilde{P}}  
\label{jorekmhdeqpert11}
\end{eqnarray}
This system allows for finite pressure and field perturbations even in absence of non--linear coupling, however to sustain a finite flow it is required to have non--linear coupling (e.g. non--ideal terms) and/or finite dynamics, as will be discussed in a next work.  

\begin{figure}
\begin{center}
\includegraphics[width=70mm]{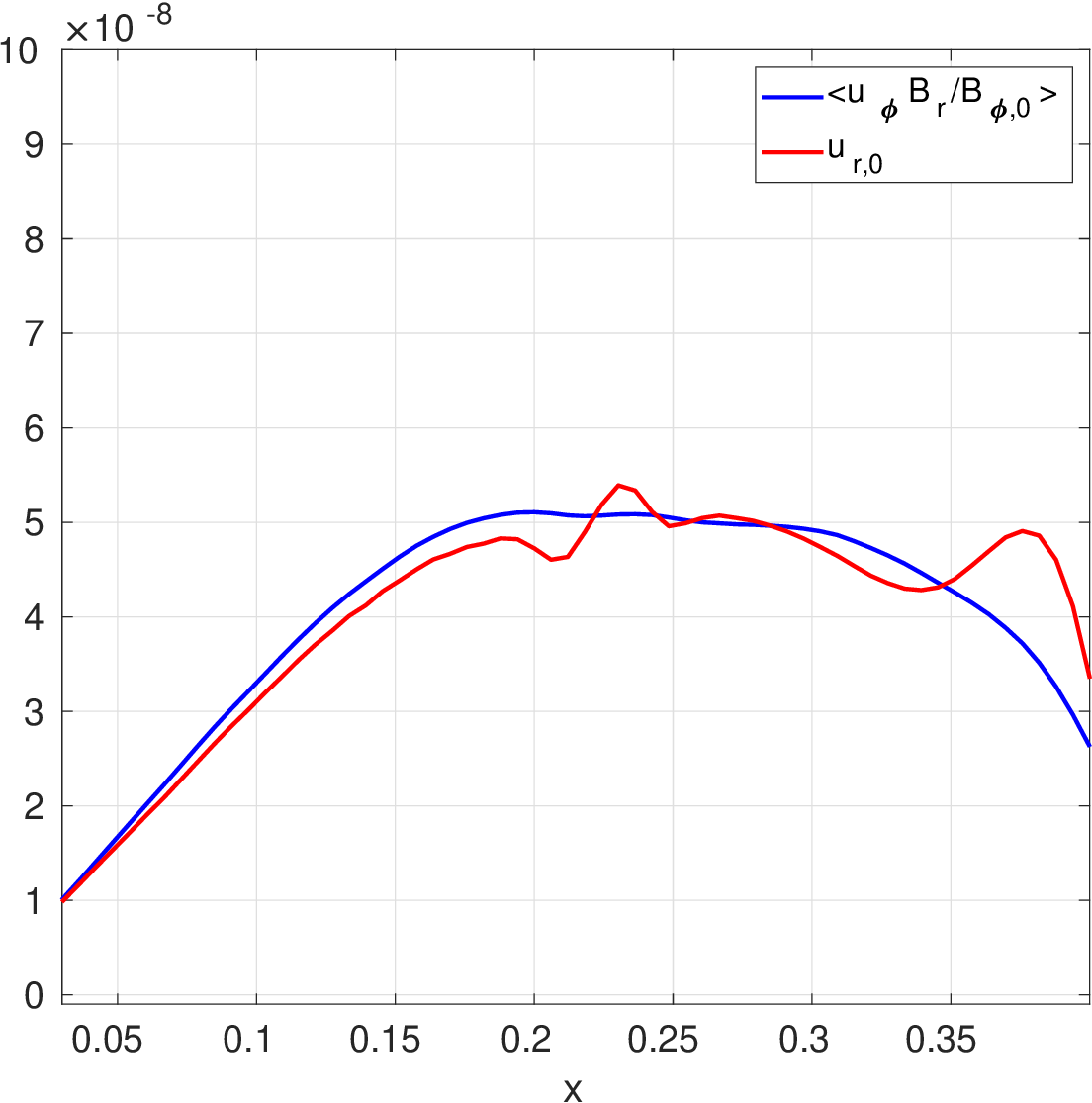}
\caption{\small Equilibrium radial flow $\ur0$ (blue line) compared to the correlation $\langle \uphit\Brhot/\Bphi0\rangle_\phi$ (red line) during the late saturated phase.}
\label{figure5}
\end{center}
\end{figure}

In figure \ref{figure5} we compare the equilibrium radial flow (flux--surface--averaged) $\langle\ur0\rangle_{f.s.a.}$ with the correlator $\langle \uphit \Brhot/\Bphi0 \rangle_{f.s.a.}$.
This comparison is meaningful only in the saturated phase since the equilibrium radial field establishes on a slower time scale than the fast linear MHD growth phase. Indeed JOREK shows that there is an equilibrium radial flow given precisely by the correlator, as predicted from equation (\ref{radfloworigin1}).

\begin{figure}
\begin{center}
\includegraphics[width=70mm]{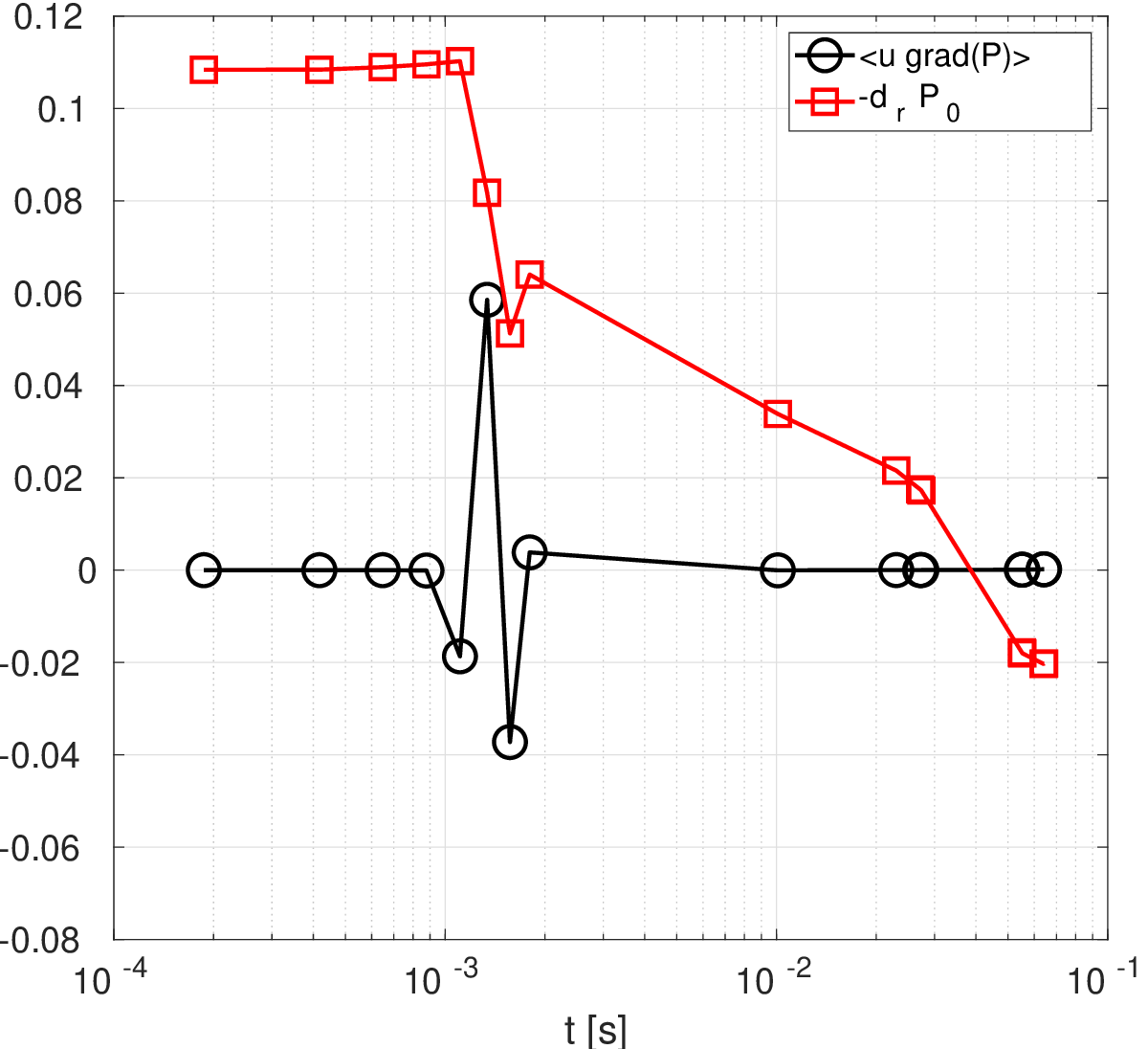}
\includegraphics[width=70mm]{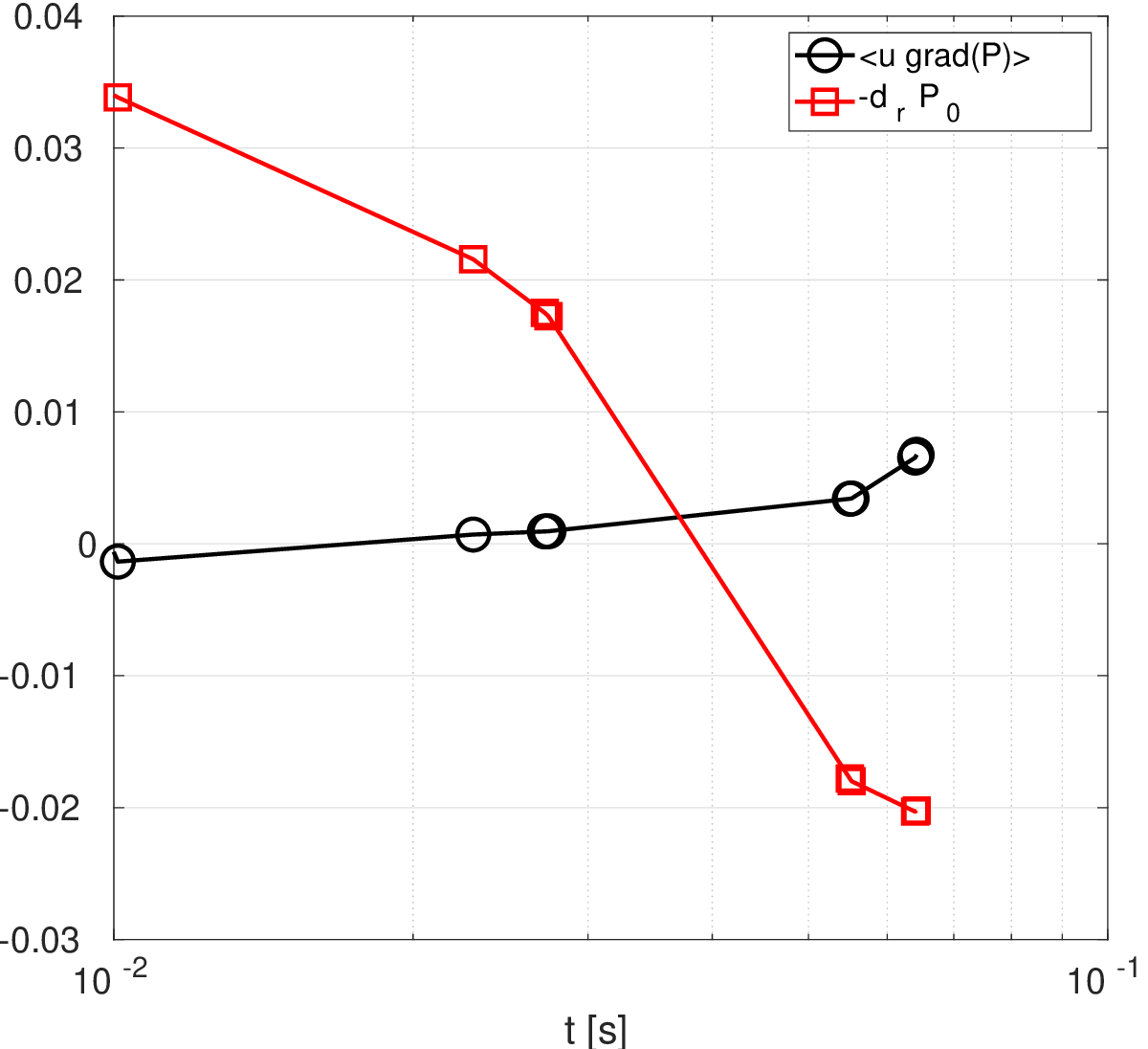}
\caption{\small Left plot: comparison between the energy flow term correlation $\langle \tilde{u}\cdot\nabla\tilde{P} \rangle$ (black circles) and the equilibrium pressure gradient (red squares), evaluated at $x\approx 0.1$ radial position. Right plot: zoom of left plot in the later phase. In both plots the correlation term is amplified artificially to make it more visible.}
\label{figure6}
\end{center}
\end{figure}

For the energy flow induced by the MHD perturbations, i.e. the first term on the right--hand--side of equation (\ref{eqpresseq1}), we show the evolution from the linear to the saturated phase together with the local pressure gradient in figure \ref{figure6}.
It can be seen that in the linear phase before $\approx 1$ ms, the term is negative, that is it is a sink of energy, as predicted above. The oscillatory phase has strong oscillations due to the radial flow perturbation strongly driven up and down as prediceted by the predator--prey model before. Finally, in the saturation phase this term is small and becomes even positive (source of energy) when the local pressure profile is hollow. 
\section{Discussion and outlook}
In this work we have presented an analytical investigation of the origin of various phenomenology linked to flux pumping, i.e. a saturated quasi--interchange mode, and compared the prediction with the results from JOREK simulations of a flux pumping typical scenario. The novel elements shown here can be summarized in the following points: 

1) a simple predator--prey model based on the evolution of kinetic and magnetic energy and the plasma pressure reproduces the simulation results including the saturation phase, giving some insights on its phenomenology. In particular, it predicts a strong relation between the saturated field amplitude and the linear growth rate, as well as the scaling of the oscillatory phase and its damping rate with the growth rate and the background transport/viscosity respectively; 

2) the $n \neq 0$ perturbations correlations analysis allows to understand the origin and behavior of the dynamo term and the energy flux term for the linear and the early saturation phase;

3) the non--ideal effects that allow the sustainment of a long--time--scale flow pattern whereas the magnetic field is quasi--stationary are given by the non--linear coupling of the different toroidal/poloidal harmonics. More in--depth analysis of this will be part of a following paper.

The comparison to the code simulations confirm the analysis results.

These findings will help in building up a reduced model of the transport induced by the saturated MHD mode. In particular, it is speculated that the long-term perturbed flow will arise from a purely dynamical state which involves the equilibrium fields. This will be proven in details in the second part to this work.
\\\\
{\bf Acknowledgments }
\\
The simulation work was partially carried out using supercomputer resources, in the Pitagora HPC at Bologna, Italy and Plasma Simulator at Rokkasho, Japan provided under the EU-JA Broader Approach collaboration and the Computational Simulation Centre of International Fusion Energy Research Centre (IFERC-CSC). The simulation work also utilized the TOK cluster at the Max Planck Institute for Plasma Physics. This work has been carried out within the framework of the EUROfusion Consortium, funded by the European Union via the Euratom Research and Training Programme (Grant Agreement No 101052200 — EUROfusion). Views and opinions expressed are however those of the author(s) only and do not necessarily reflect those of the European Union or the European Commission. Neither the European Union nor the European Commission can be held responsible for them.

\section*{References}
\bibliographystyle{iopart-num}
\bibliography{references_list}
\end{document}